\documentclass[manuscript]{aastex}

\shorttitle{Potassium and Lithium in comets C/2011 L4 and C/1965 S1}
\shortauthors{Fulle et al.}

\begin{document}


\title{Potassium detection and Lithium depletion in comets
       C/2011 L4 (Panstarrs) and C/1965 S1 (Ikeya-Seki)}


\author{M. Fulle\altaffilmark{1}}
\affil{INAF - Osservatorio Astronomico,
    Via Tiepolo 11, I-34143 Trieste Italy}
\email{fulle@oats.inaf.it}

\author{P. Molaro\altaffilmark{1}}
\affil{INAF - Osservatorio Astronomico,
    Via Tiepolo 11, I-34143 Trieste Italy}

\author{L. Buzzi and P. Valisa\altaffilmark{2}}
\affil{Societ\`a Astronomica Schiaparelli,
    Via Beato Angelico 1, Varese Italy}



\begin{abstract}
On 21 March 2013 high-resolution slit spectrographs of comet C/2011 L4 (Panstarrs),
at the heliocentric distance $r = 0.46$ Astronomical Units (AU), were obtained at the
Osservatorio Astronomico Campo dei Fiori, Italy. Emission lines of sodium were the
strongest in the spectrum as is common in comets, but also potassium lines were detected.
These have been rarely observed in comets since the apparition of brightest comet
C/1965 S1 (Ikeya-Seki). Lithium was not detected, and stringent upper limits of its
abundance compared to other alkali were derived. We obtain abundance ratios Na/K =
$54 \pm 14$ and Na/Li $\ge 8 ~10^3$. As well as in Mercury exosphere \citep{leb11}, we
show that photoionization at the beginning of the alkali tails may increase the solar
ratio Na/K = 15.5 \citep{asp09} by a factor 3, close to that required to match the observed
value. In the same tail position the Na/Li ratio increases by a factor 2 only, very far
from the factor $\ge 8$ required to match an original meteoritic ratio. We apply the same
model to similar alkali data \citep{pre67} of comet C/1965 S1 (Ikeya-Seki), obtaining
consistent results. An original solar Na/K ratio fits the observed value at the beginning
of the alkali tails within the slit size, whereas Li is depleted by a factor $\ge 8$.
\end{abstract}





\section{Introduction}

Comets provide unique information on the cosmic abundances of the solar nebula which
collapsed to form the Solar System. Besides sodium, already detected in many comets,
very few data regard the alkali content of comet nuclei. In particular, potassium was
remotely detected \citep{pre67} in spectra of comet C/1965 S1 (Ikeya-Seki), where the
Na/K ratio was as high as that observed in Mercury \citep{kil10}. A detection of the
potassium line at 7698.9645 \AA ~in comet C/1995 O1 (Hale-Bopp) was reported by
\cite{fit97}, but no Na/K ratio has been extracted from this spectrum yet. Laboratory
analyses of samples collected at comet 81P/Wild 2 showed that most potassium is in
form of eifelite and K-feldspar grains \citep{zol06}. Detections of the potassium and
lithium lines were also reported from the impacts of comet D/1993 F2 (Shoemaker-Levy
9) on Jupiter \citep{roo95}. The Na/Li ratio extracted from spectra of the plume in
Jupiter's atmosphere were consistent with a meteoritic ratio \citep{cos97}, although
these transient emissions, unlike resonant fluorescence, could not be readily converted
into atomic abundances and were probably contaminated by the alkali of Jupiter's deep
atmosphere. In this report we discuss the remote detection of potassium in comet C/2011
L4 (Panstarrs), which allows us to discuss the Na/K abundance in this comet. As well as
in models of Mercury's exosphere \citep{leb11}, we will discuss the transfer of alkali
atoms from the parent bodies (mainly the dust grains in the sunward coma) to the
beginning of alkali tails \citep{ful07} where they have been observed. This will allow
us to face actual estimates of lithium abundances in comets. The physical process
relating alkali line intensities to atomic abundance is very simple, namely resonance
fluorescence \citep{swi41}. The absorption of solar radiation in the resonance
transitions populates atomic upper levels, which trickling down give rise to the
emission lines. The population in the upper level depends on the energy available at
the considered wavelength, i.e. upon whether a Fraunhofer line comes in the way of
absorption or not. These Fraunhofer lines as seen from the comet have different Doppler
shifts depending on the radial heliocentric velocity $v$ (the so-called Swing effect).

\section{The observations of comet C/2011 L4 (Panstarrs)}


High resolution ($\lambda/\Delta\lambda \approx 10^4$ in the spectral range 424-864 nm)
Echelle spectra (Fig. 1) of comet C/2011 L4 (Panstarrs) were obtained on 21.8 UT March
2013 (mid exposure time) with the Multi-Mode Spectrograph mounted on the 0.6m telescope at
Osservatorio Astronomico Schiaparelli located in Campo dei Fiori, Varese, Italy \citep{ash12}.
The multi-order Echelle spectra were absolutely flux calibrated against spectra of the
standard star HR 464 located on the sky nearby, observed immediately after the comet, and
then merged into a 1D continuous spectra. The slit was E-W oriented and guided on the
brightest coma by means of a monitor of a guiding ccd-camera covering $3 \times 4$ arcmin$^2$
on the sky. The slit width projected on the sky was set to 4 arcsec, and the slit length to
17 arcsec. The total exposure time was 40 min. At the observations, the Sun-comet distance
was $r = 0.46$ AU; the Earth-comet distance $\Delta = 1.19$ AU; the comet was receding from
the Sun at a speed of 36 km s$^{-1}$, and from Earth at a speed of 14 km s$^{-1}$. The
Sun-comet-Earth phase angle was 54 degrees. Images of the same dust coma showed that the
apex distance (i.e. the distance between the brightest inner coma and its sunward boundary)
was about 100 arcsec, corresponding to $10^5$ km projected along the Sun-comet vector.

The cometary spectrum show a number of emission features with very prominent NaI but also
with the KI lines clearly detected. Other identified emissions include the C$_2$ Swan band
dv=0 and the satellite dv=1 e dv=-1, NH2 (7-0) and NH2 (9-0), and the [OI] at $\lambda\lambda$
6300 and 6363 \AA ~redshifted of 14 km s$^{-1}$ (Fig. 1). The alkali lines are shown in Figs.
2 and 3, where the reconstructed profile is estimated after accounting for the contamination
of the telluric O$_2$ absorption bands and of the solar spectrum reflected by the cometary dust.
These corrections are computed by means of a twilight spectrum which was also recorded with the
same spectrograph setup. The KI line at $\lambda\lambda$ 7664.8991 \AA ~(binned thick black line
in the left panel of Fig. 2) is blended with two telluric O$_2$ lines at 7664.73 and 7665.79
\AA, and with the KI solar absorption produced by comet dust reflection of the solar light
and redshifted by 50 km s$^{-1}$. In order to reconstruct the cometary KI emission, we used
a twilight spectrum (continuous red line in the same panel) to which we subtracted the solar
KI absorption at rest (left dashed red line in the same panel) and added the KI absorption
redshifted by 50 km s$^{-1}$ (right dashed red line in the same panel). The obtained local
continuum (dashed blue line in the same panel) well reproduces the wings of the absorption.
The resulting emission KI line (continuous blue line in the same panel) is the difference
between the original spectrum (binned thick black line in the same panel) and the reconstructed
local continuum (dashed blue line in the left panel of Fig. 2). The other KI emission line at
$\lambda\lambda$ 7698.9645 \AA ~(right panel of Fig. 2) and the NaI lines at $\lambda\lambda$
5895.9242 and 5889.9510 \AA ~(Fig. 3) are not contaminated by telluric bands, but are partially
eroded by the corresponding solar absorption line reflected by the comet dust. In these cases,
the twilight feature (continuous red line) has been redshifted by 50 km s$^{-1}$ to match the
red wing and to reconstruct the true emission profile (continuous blue line in Figs. 2 and 3).
The intensities of the corrected alkali lines are reported in Table 1. No emission is detected at
the position of the LiI $\lambda\lambda$ 6707.78 \AA~ line (Fig. 4) and the 3$\sigma$ upper limit
for the LiI emission is shown in Table 1.

The abundance Na/x of sodium related to an atom x (Table 1) is extracted from the line intensity $I$
(Table 1) by means of the relationship Na/x = ($g_x/g_{Na}) (I_{Na}/I_x$), where the $g$-factors $g_x$
at 1 AU are computed (Table 1) as a function of the heliocentric radial velocity $v$ (thus taking into
account the Swing effect) using the high resolution visible solar flux \citep{kur84} and the oscillator
strengths for the observed resonant lines \citep{mor03,mor04}. Regarding $g_K$ at $\lambda$ = 7664.8991
\AA, in the range $35 \le v \le 45$ km s$^{-1}$ the solar spectrum is dominated by the strong absorption
of telluric oxigen, so that we had to assume the same mean value obtained at smaller and larger
$v$-values. The intensity ratio between the two Na and K lines ($1.7 \pm 0.1$ for Na and $1.6 \pm 0.3$
for K, respectively) matches the corresponding $g$-factor ratio ($2.0 \pm 0.4$ for Na and $1.9 \pm 0.1$
for K, respectively), so that we can exclude significant optical thickness in the lines.

\section{Model of alkali tails of comet C/2011 L4 (Panstarrs)}

Five processes are expected to extract alkali atoms from the parent body, namely (i)
thermal desorption, (ii) photon-stimulated desorption, (iii) solar wind sputtering,
(iv) micro-meteoroid vaporization, and (v) photodissociation of parent molecules.
Given the atomic parameters of alkali and the results of laboratory experiments on cosmic
analogues, it is expected that the cloud of atoms leaving the parent surface maintains
its original (presumably solar) abundance \citep{leb11}. Hereinafter, when we refer to
solar lithium abundance, we mean that measured in meteorites \citep{asp09}. After the
atom release, many phenomena have to be taken into account to infer the Na/K ratio in
Mercury's exosphere \citep{leb11}, the main one occurring in comets too is photoionization
by solar UV radiation. All alkali atoms are pushed in the anti-sunward direction by solar
radiation pressure, with accelerations $a = G M_\odot (\sum_\lambda \beta_{\lambda}) ~r^{-2}$
in the comet reference frame. Here $G$ is the gravitational constant, $M_\odot$ is the
Sun mass, $r$ is the Sun-comet distance, and $\beta$ is the ratio between solar gravity
and radiation pressure forces. From the $g$-factors we compute (Table 1)
$\beta_{\lambda} = g_\lambda ~(1 AU)^2 / (c ~m ~G M_\odot)$, where $c$ is the velocity
of light and $m$ is the mass of the atom \citep{ful04}. As it happens for the ejection of
iron atoms \citep{ful07}, alkali atoms should be mainly ejected from dust, which should be
mostly ejected from the comet nucleus into the sunward sector of the coma. Before reaching
the observation slit, the atoms have to cover at most $10^5$ km (i.e. the apex distance) in
the anti-sunward direction, which requires the flight times $\Delta t_{Na} = 9.5 ~10^3$ s and
$\Delta t_K = 1.2 ~10^4$ s, respectively. Taking into account the photoionization lifetimes
listed in Table 1, the original Na/K ratio would increase by a factor 3, so that we cannot
exclude an original solar ratio Na/K = 15.5 \citep{asp09}. In other words, what we actually
observed was the beginning of the alkali tail, where K is depleted vs sodium by its shorter
photoionization lifetime. In order to cover the same distance of $10^5$ km, lithium atoms
need a flight time $\Delta t_{Li} = 4.1 ~10^3$ s. Taking into account its photoionization
lifetime (Table 1), we get an increase of the original Na/Li ratio by a factor 2 only, very
far from the factor 8 required to match the solar ratio Na/Li = $10^3$ \citep{asp09} to the
observed one. The required factor 8 would be observed at $7 ~10^5$ km from the nucleus,
corresponding to 12 arcmin, a distance impossible to accept since the slit tracking was
done on the brightest inner coma with a mean diameter of 3 arcmin. We conclude that lithium
in comet C/2011 L4 is depleted by a factor $\ge 4$ with respect to the solar ratio.

\section{Model of alkali tails of comet C/1965 S1 (Ikeya-Seki)}

Sodium and potassium were detected in spectra of comet C/1965 S1 (Ikeya-Seki) at the distance
$r = 0.14$ AU, when the comet was receding from the Sun at a velocity of 110 km s$^{-1}$
\citep{pre67}. We consider the observed intensities of the alkali lines Na $\lambda\lambda$
5889.9510 and KI $\lambda\lambda$ 7698.9645 measured by \cite{pre67}, $I_{Na}/I_K = 80$ and
$I_{Na}/I_{Li} \ge 1.3 ~10^4$, respectively. By means of the $g$-factors reported in Table 1,
we obtain Na/K = $50 \pm 11$ and Na/Li $\ge 3.3 ~10^4$, that are values in close agreement
with those measured in C/2011 L4. \cite{pre67} obtained even higher values assuming excitation
mechanisms more complex than fluorescence. However, the resulting excitation temperatures are
quite different among different atoms, and their physical interpretation is difficult. Due to
the intensity of sodium lines, these were saturated in the obtained photographic spectra, so
that we cannot exclude that the sodium lines were optically thick. However, since sodium is
by far the most abundant among alkali atoms, we can assume that optical thickness affected
potassium (and lithium if any) much less than sodium. This would further increase the ratios
listed above. We assume that the alkali atoms have to cover an anti-sunward distance of $4 ~10^4$
km (exactly matching the slit length of one arcmin used in the spectrograph setup) after they
are ejected by the dust in the sunward coma. In order to cover such a distance, the atoms need
flight times $\Delta t_{Na} = 1.9 ~10^3$ s and $\Delta t_K = 2.2 ~10^3$ s, respectively. Taking
into account the photoionization lifetimes listed in Table 1, the original Na/K ratio would
increase by a factor 8, more than required to match an original solar Na/K ratio. Therefore, an
original solar Na/K ratio would remain consistent with observations even if the Na/K ratio in
the alkali tail had been a factor 3 higher due to optically thick sodium lines. On the other
hand, lithium atoms need a flight time $\Delta t_{Li} = 7.8 ~10^2$ s. Taking into account its
photoionization lifetime, we get an increase of the original Na/Li ratio by a factor 4 only,
very far from the factor 33 required to match the solar Na/Li ratio to the observed one. We
conclude that lithium was depleted with respect to the solar abundance in comet C/1965 S1 too,
by a factor $\ge 8$, even more constraining than in comet C/2011 L4. Actual detection of lithium
in future bright comets may help to understand if such a depletion is original in comet nuclei,
or is due to a less efficient process of extraction with respect to other alkali atoms.



\acknowledgments

We thank Jacques Crovisier for his comments which significantly improved the original manuscript,
and Pierluigi Selvelli for useful discussion on atomic emission processes. Correspondence and
requests for material should be addressed to M. Fulle (fulle@oats.inaf.it).


\clearpage



\clearpage

\begin{figure}
\epsscale{1.}
\plotone{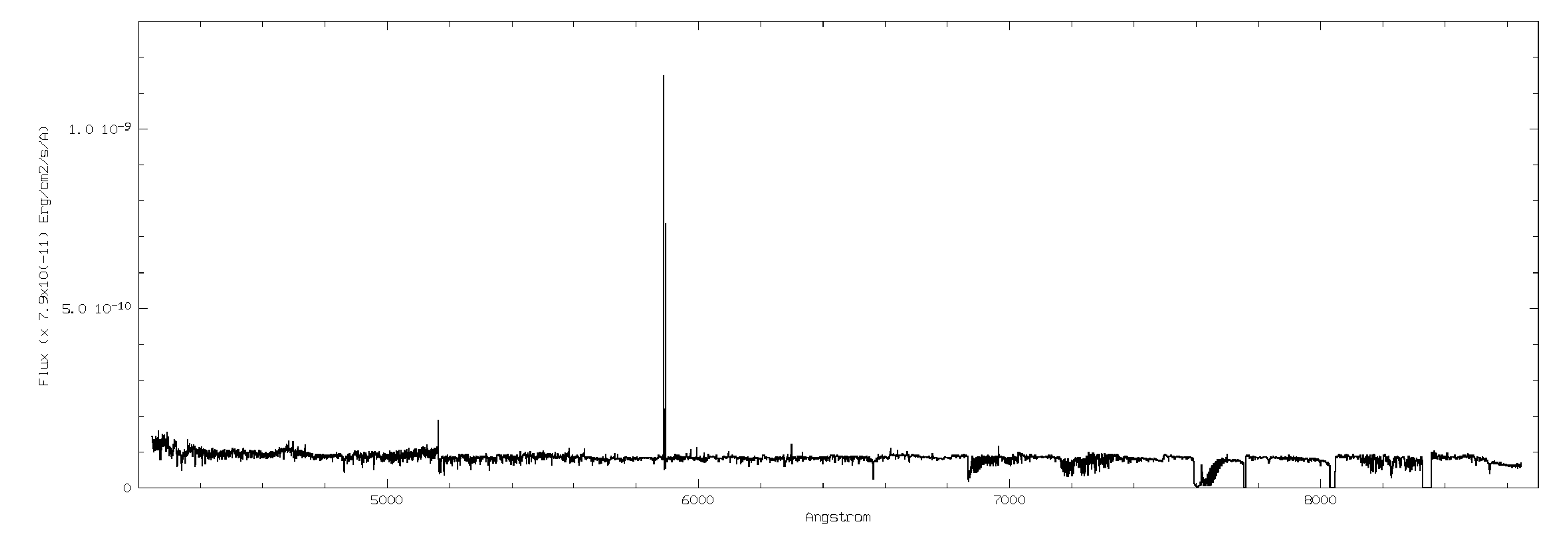}
\caption{The C/2011 L4 spectrum shown over the complete observed wavelength range. It is
dominated by the sodium emission and by O$_2$ telluric absorption bands.\label{fig1}}
\end{figure}

\clearpage

\begin{figure}
\epsscale{1.}
\plotone{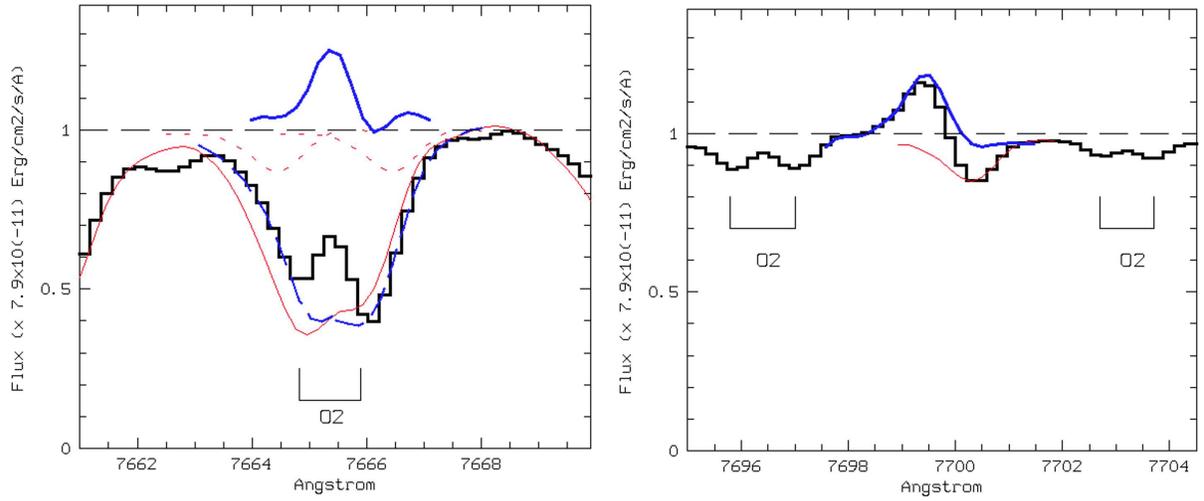}
\caption{The C/2011 L4 spectrum is shown in the region of the KI $\lambda\lambda$ 7664.8991
\AA ~(left panel) and in the region of the KI $\lambda\lambda$ 7698.9645 \AA ~(right panel).
Binned thick black line: observed spectrum. Continuous blue line: reconstructed spectrum
after the correction of telluric O$_2$ absorption bands and solar continuum reflected by
cometary dust (see Section 2 for the details of the procedure).\label{fig2}}
\end{figure}

\clearpage

\begin{figure}
\epsscale{1.}
\plotone{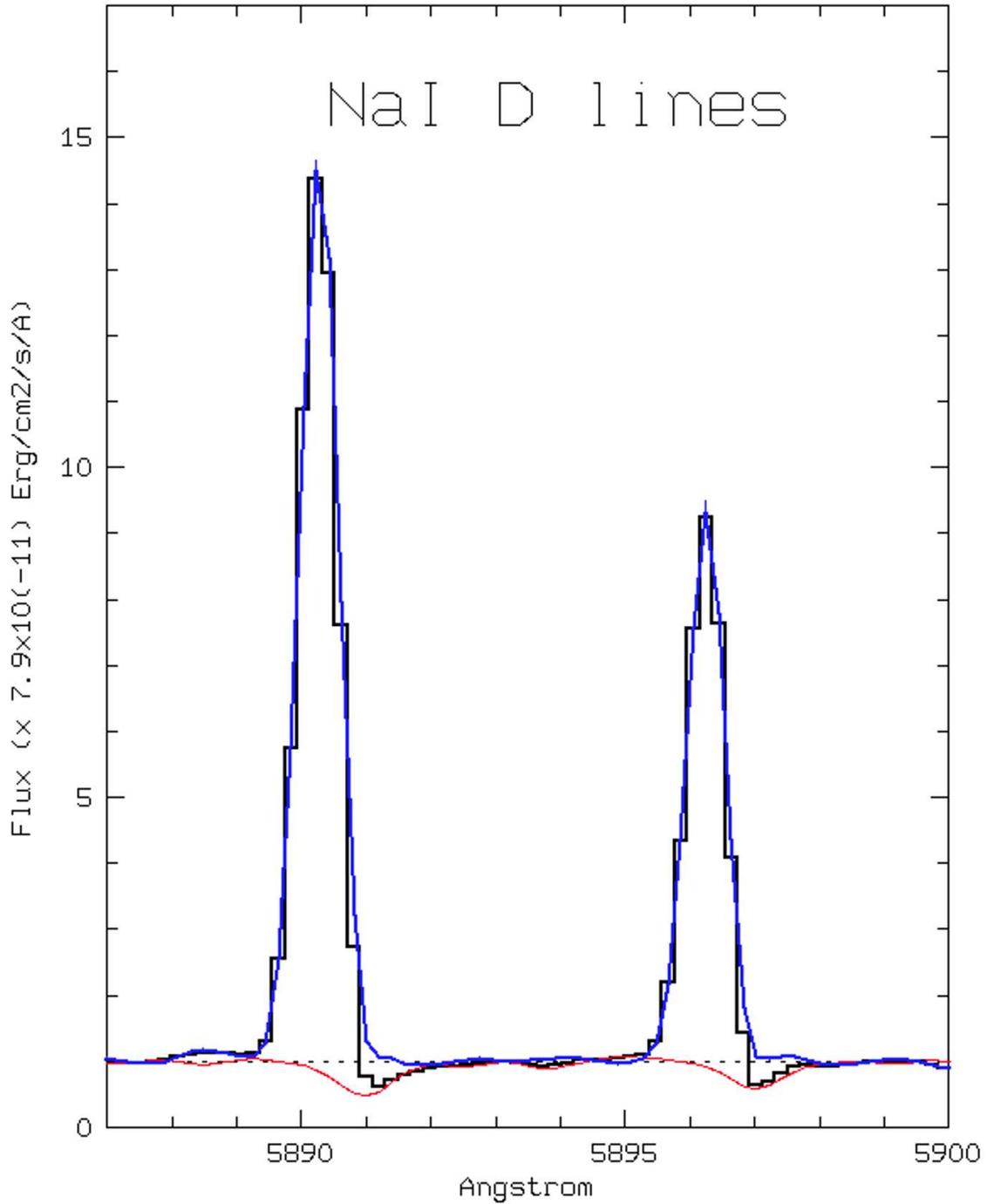}
\caption{The C/2011 L4 spectrum is shown in the region of the NaI D lines at $\lambda\lambda$
5895.9242 and 5889.9510 \AA ~(binned thick black line). The red wing of cometary emissions
are partially eroded by the corresponding solar NaI feature reflected by the comet dust. The
twilight feature (red continuous line) has been redshifted by 50 km s$^{-1}$ to match the red
wing and to reconstruct the true emission profiles (continuous blue line).\label{fig3}}
\end{figure}

\clearpage

\begin{figure}
\epsscale{1.}
\plotone{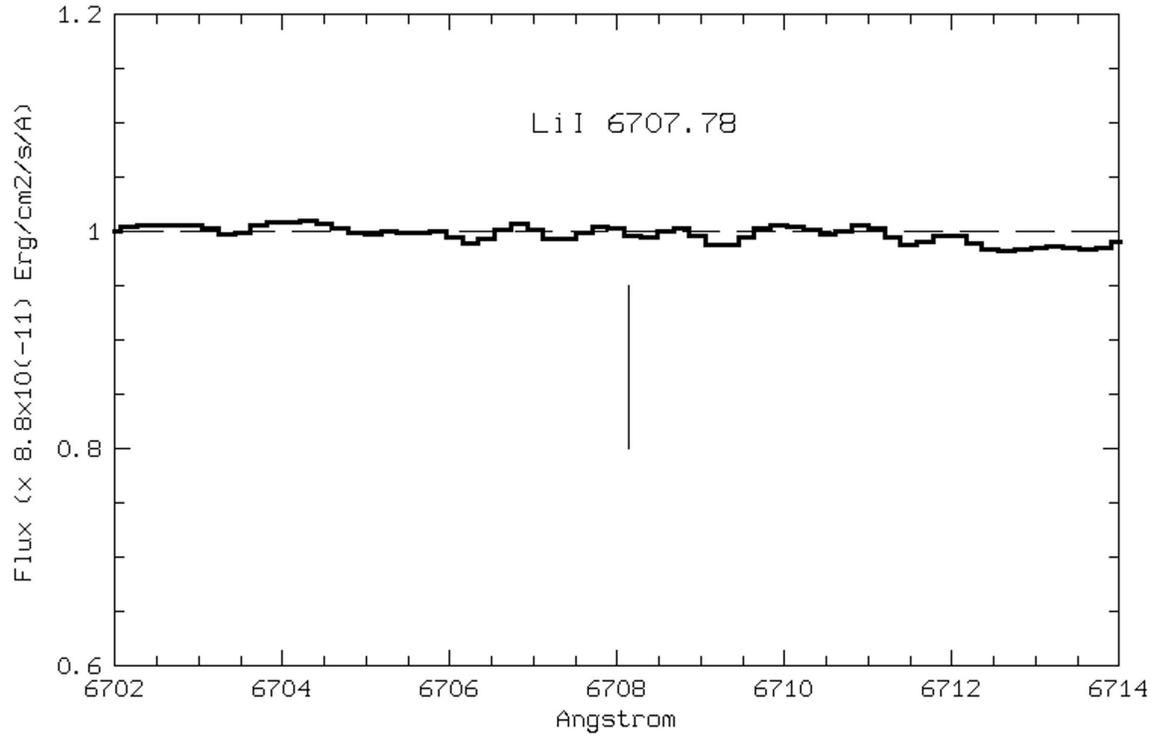}
\caption{The C/2011 L4 spectrum is shown in the region of the LiI line at $\lambda\lambda$
6707.78 \AA ~(binned thick black line) after subtraction of the contamination of the solar
spectrum. No emission was detected at the position of lithium line redshifted by 14 km
s$^{-1}$ (vertical line).\label{fig4}}
\end{figure}

\clearpage

\begin{deluxetable}{ccccccccccc}
\tabletypesize{\scriptsize}
\tablecaption{Atomic parameters of alkali for heliocentric velocity $35 \le v \le 140$ km s$^{-1}$.\label{tbl-1}}
\tablewidth{0pt}
\tablehead{
   \colhead{Atom} & \colhead{$\lambda$\tablenotemark{a}} & \colhead{$I$\tablenotemark{b}} & \colhead{$g$\tablenotemark{c}}
   & \colhead{Na/x\tablenotemark{d}} & \colhead{Na/x\tablenotemark{e}} & \colhead{$\beta_\lambda$} & \colhead{a\tablenotemark{f}}
   & \colhead{a\tablenotemark{g}} & \colhead{$\tau$\tablenotemark{h}} & \colhead{$\tau$\tablenotemark{j}}
}
\startdata
Li &6707.78 &$\le 0.027$   &$9.15 \pm 0.15$ &$\ge 8 ~10^3$ &$\ge 4 ~10^3$ &$440 \pm 7$  &$12.3 \pm 0.2$ &$133 \pm 2$
   &4 ~10$^3$                     &4 ~10$^2$\tablenotemark{k}    \\
Na &5889.95 &$85 \pm 1$    &$3.23 \pm 0.34$ &1             &1             &$47 \pm 5$   &$2.0 \pm 0.2$  &$22 \pm 2$
   &4.0 ~10$^4$\tablenotemark{m}  &3.7 ~10$^3$                   \\
Na &5895.92 &$49 \pm 1$    &$1.62 \pm 0.17$ &\nodata       &\nodata       &$24 \pm 3$   &\nodata        &\nodata
   &\nodata                       &\nodata                       \\
K  &7664.90 &$2.1 \pm 0.2$ &$4.12 \pm 0.21$ &$54 \pm 14$   &$18 \pm 5$    &$35 \pm 2$   &$1.5 \pm 0.1$  &$16 \pm 1$
   &9.1 ~10$^3$\tablenotemark{n}  &8.4 ~10$^2$                   \\
K  &7698.96 &$1.3 \pm 0.1$ &$2.16 \pm 0.03$ &\nodata       &\nodata       &$18 \pm 0.2$ &\nodata        &\nodata
   &\nodata                       &\nodata                       \\
\enddata
\tablenotetext{a}{wavelength of the emission line (\AA)}
\tablenotetext{b}{measured intensity of the emission line ($10^{-11}$ erg s$^{-1}$ cm$^{-2}$ \AA$^{-1}$)}
\tablenotetext{c}{computed $g$-factor ($10^{-11}$ erg s$^{-1}$)}
\tablenotetext{d}{relative abundance at the source withouth photoionization}
\tablenotetext{e}{relative abundance at the source with photoionization observed at $10^5$ km from the source}
\tablenotetext{f}{anti-sunward acceleration (m s$^{-2}$) at r = 0.46 AU}
\tablenotetext{g}{anti-sunward acceleration (m s$^{-2}$) at r = 0.14 AU}
\tablenotetext{h}{photoionization lifetime (s) at r = 0.46 AU}
\tablenotetext{j}{photoionization lifetime (s) at r = 0.14 AU}
\tablenotetext{k}{\cite{pre67}}
\tablenotetext{m}{\cite{ful07}}
\tablenotetext{n}{\cite{hue92}}
\end{deluxetable}

\end{document}